\documentclass{aac}
\usepackage[numbers]{natbib}
\usepackage{rotating}
\usepackage{graphicx}
\usepackage[dvipsnames]{xcolor}
\usepackage{siunitx}
\usepackage{bm}

\begin{document}

\title{Modeling Betatron Radiation Diagnostics for E-310 - Trojan Horse}

\author[aff1,aff2,aff3]{Monika Yadav\corref{cor1}}
\author[aff1]{Claire Hansel}
\author[aff1]{Yumeng Zhuang}
\author[aff1]{Brian Naranjo}
\author[aff1]{Nathan Majernik}
\author[aff2,aff3]{Aravinda Perera}
\author[aff1]{Yusuke Sakai}
\author[aff1]{Gerard Andonian}
\author[aff1]{Oliver Williams}
\author[aff1]{Pratik Manwani}
\author[aff3,aff5]{Javier Resta-Lopez}
\author[aff2,aff3]{Oznur Apsimon}
\author[aff2,aff3]{Carsten Welsch}
\author[aff2,aff4]{Bernhard Hidding}
\author[aff1]{James Rosenzweig}

\affil[aff1]{Department of Physics and Astronomy, University of California Los Angeles, California 90095, USA}
\affil[aff2]{Cockcroft Institute, Warrington WA4 4AD, UK}
\affil[aff3]{Department of Physics, University of Liverpool, Liverpool L69 3BX, UK}
\affil[aff4]{Department of Physics, University of Strathclyde, Glasgow G1 1XQ, UK}
\affil[aff5]{ICMUV, Instituto de Ciencia de Materiales, Universidad de Valencia, 46071 Valencia, Spain}
\corresp[cor1]{monika.yadav@liverpool.ac.uk}

\maketitle

\begin{abstract}
The E-310 experiment at the Facility for Advanced Accelerator Experimental Tests II (FACET-II) at SLAC National Accelerator Laboratory aims to demonstrate the creation of high brightness beams from a plasma photocathode. Betatron radiation will be measured by a Compton spectrometer, currently under development at UCLA, to provide single-shot, nondestructive beam diagnostics. We give a brief overview of this spectrometer as well as double differential spectrum reconstruction from the spectrometer image and beam parameter reconstruction from this double differential spectrum. We discuss three models for betatron radiation: an idealized particle tracking code which computes radiation from Liénard-Wiechert potentials, a quasi-static particle-in-cell (PIC) code which computes radiation from Liénard-Wiechert potentials, and a full PIC code which computes radiation using a Monte Carlo QED method. Spectra computed by the three models for a simple case are compared. 
\end{abstract}

\section{INTRODUCTION}
Plasma wakefield accelerators (PWFAs) provide unique opportunities for the generation of high quality, short-pulse electrons beams which are an ideal basis for high quality radiation generation. For an underdense plasma, where the plasma density is less than the beam density, plasma electrons are blown out to approximately the plasma skin depth, $k_p^{-1} = c/\omega_p$, where $\omega_p$ is the plasma frequency. Within the bubble formed by this blowout, the uniformly distributed plasma ions generate strong, linear, transverse focusing fields \cite{JBR}. The betatron motion of beam electrons in these fields generates characteristic X-ray radiation. Betatron radiation from plasma accelerators can be used to create novel compact X-ray light sources \cite{Sebastian_2013} as well as for single-shot nondestructive diagnostics of beam-plasma interactions.

The planned E-310 experiment at FACET-II aims to generate high brightness beams using Trojan Horse plasma photocathode, building off the previous E-210 experiment at FACET which was a proof of concept to demonstrate beam creation \cite{E210}. Betatron radiation will be used to provide diagnostics for this experiment, as well as other plasma acceleration experiments at FACET-II. In this paper, we discuss the novel spectrometer which will be used to measure this radiation, numerical models of betatron radiation applicable to both light sources and beam diagnostics, and algorithms for the reconstruction of both the radiation spectrum and the beam parameters from the spectrometer output.

\section{UCLA SPECTROMETER AND RADIATION SPECTRUM RECONSTRUCTION}
\label{sec:Reconstruction}

\begin{figure}[h!]
    \centering
    \includegraphics[width=70mm,scale=0.5]{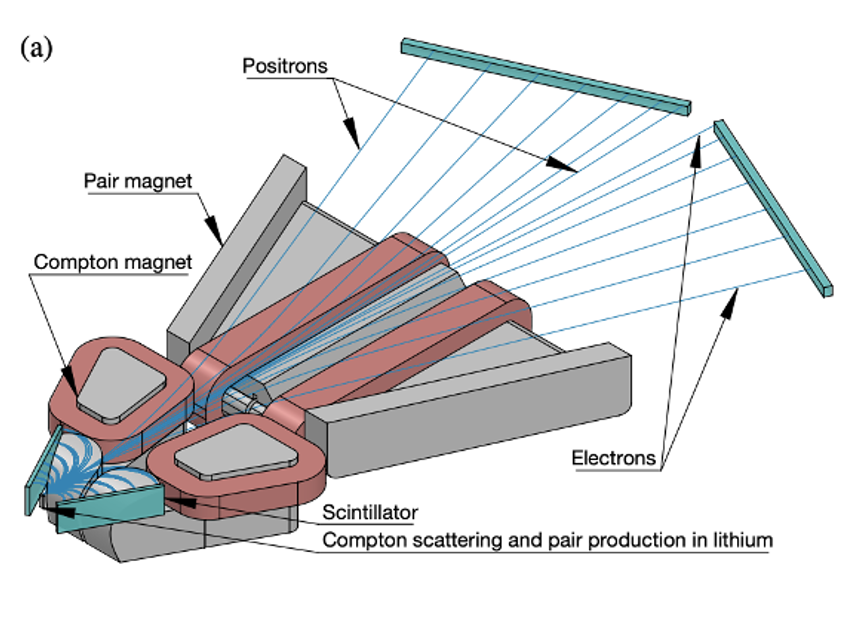}
    \label{fig:uhbemittance}
    \centering
    \includegraphics[width=70mm,scale=0.5]{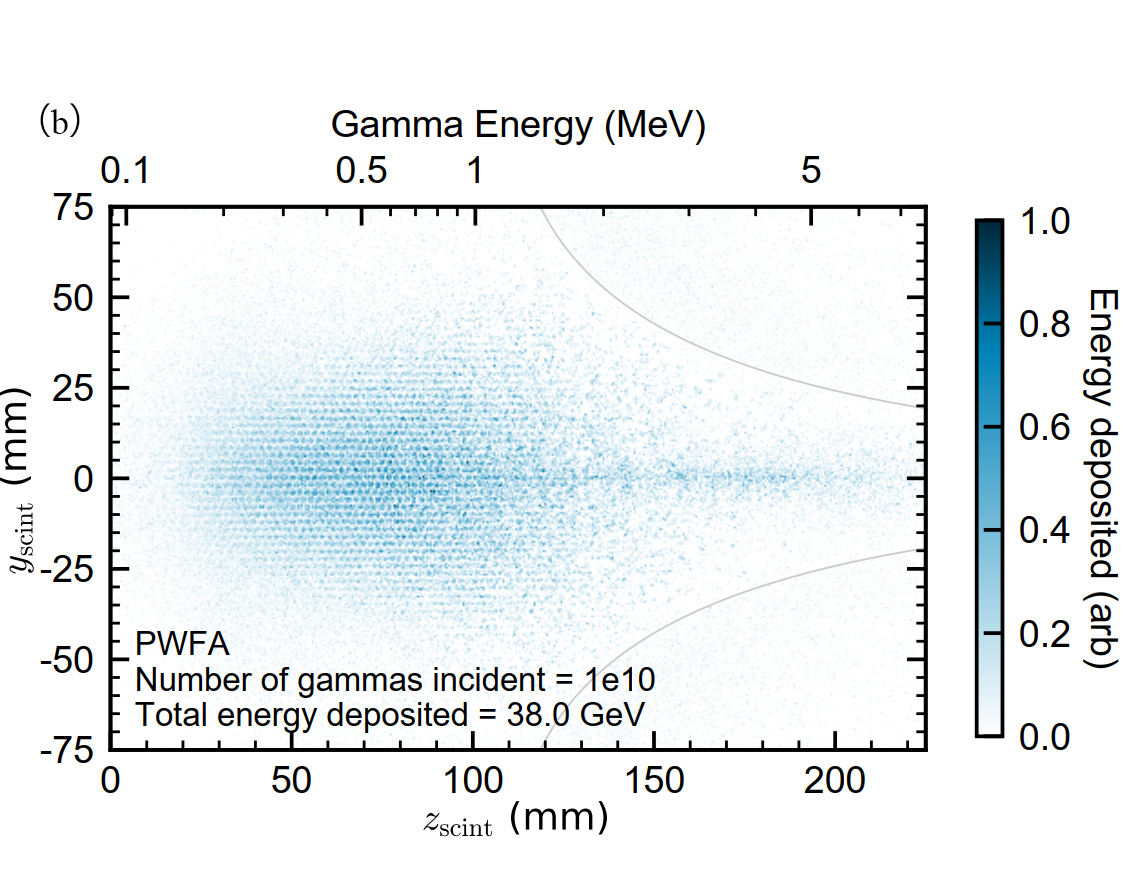}
    \caption{\textbf{(a)} Diagram of UCLA spectrometer.  \textbf{(b)} Simulated image on the Compton scintillator produced by a Geant4 model of the spectrometer.
}
    \label{fig:spectrometer}
\end{figure}

A novel spectrometer, shown schematically in Fig.~\ref{fig:spectrometer}, is currently under development at UCLA. This spectrometer is designed to measure double differential photon spectra over several decades of energy: from tens of keV through ten GeV, with an angular resolution of $\SI{100}{\mu rad}$. The emitted radiation enters the spectrometer from the lower left of Fig.~\ref{fig:spectrometer}(a) and interacts with a lithium converter, causing Compton scattering and pair production. The lower energy photons ($\SI{313}{keV}$ to $\SI{20}{MeV}$) interact primarily via Compton scattering and the resulting electrons are bent by the Compton magnet into the two scintillators on the lower left. Higher energy ($\SI{125}{MeV}$ to $\SI{8}{GeV}$) photons primarily exhibit pair production with the resulting electrons and positrons continuing past the Compton magnet, before being further curved by the pair magnet, and finally interacting with the two scintillators on the top right.

Both an expectation-maximization, (EM)-based, and a machine-learning, (ML)-based, algorithm have been used to reconstruct the double differential spectrum from the scintillator values. The reconstruction problem is as follows: given $n(d)$, the energy deposited in the finite scintillator bin $d$, and $p(b, d)$, the energy deposition in scintillator bin $d$ of photons coming from energy bin $b$, what must be obtained is $\lambda(b)$,  the total energy deposited by the gamma photons of each energy per bin. In other words, given the overall energy deposition pattern on the scintillator, and a basis composed of the deposition pattern for each possible binned energy of incoming photons, we are required to find the true total binned energy spectrum of the incoming photons. This is achieved using two approaches:

\begin{itemize}
\item \textit{EM algorithm} -- The EM  algorithm finds a set of parameters that maximizes the likelihood of a given outcome under a set of constraints, by iteration. We adapted the approach introduced by Vardi and Shepp \cite{Vardi}.

\item \textit{Fully connected neural network} -- A fully connected neural network is a series of matrix multiplications plus bias terms with activation functions applied on each level. All the weights in the matrices and biases are updated in the training process in the direction that minimizes the error of the output.
\end{itemize}

\begin{figure}[h!]
    \centering
    \includegraphics[width=140mm,scale=0.5]{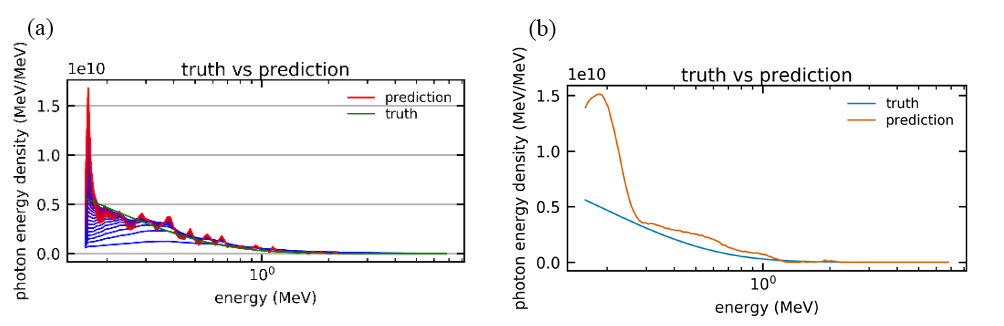}
    \caption{\textbf{(a)} Reconstructed PWFA spectrum by EM algorithm. Blue curves are early solutions which converge to red curve via the iteration process. 
    \textbf{(b)} Reconstructed PWFA spectrum by a fully-connected neural network.}
    \label{fig:reconstruction}
\end{figure}

To benchmark these approaches, Geant4 \cite{geant} is used to compute the scintillator values from a given radiation spectrum. Example reconstruction results for both algorithms are shown in Fig.~\ref{fig:reconstruction}, shown with the ground truth spectrum used to generate the signal. The agreement is generally good, but reconstruction at low photon energies is challenging due to the spread in Compton scattering angle and the resulting weak and ambiguous signal, but both methods give satisfactory results for 1D on-axis energy spectra above 200 keV. 

\section{BEAM PARAMETER RECONSTRUCTION FROM BETATRON RADIATION}
In plasma wakefield accelerators operating in the blowout regime, linear focusing fields inside the bubble cause beam electrons to undergo simple harmonic betatron oscillations with angular wavenumber, $k_{\beta} = \sqrt{{2 \pi r_e Z_i n_0}/{\gamma}}$, where $r_e$ is the classical electron radius, $Z_i$ is the ion charge state, and $n_0$ is the plasma density. Each electron generates radiation with an equivalent undulator parameter, $K = \gamma k_{\beta} x_{\beta}$, where $x_{\beta}$ is the amplitude of the betatron oscillation. This radiation is emitted within a cone of angle approximately $K / \gamma$. There are three regimes of this radiation: for $K \ll 1$, weak undulator radiation, monochromatic radiation is emitted with a single sharp peak in the spectrum; for $K \approx 1$, strong undulator radiation, the spectrum also includes higher harmonics of the fundamental wavelength; for $K \gg 1$, wiggler radiation, the radiation tends towards a smooth, synchrotron spectrum characterized by a critical photon energy of $\epsilon_c = (3/2) \hbar c K \gamma^2 k_{\beta}$.  For the parameters of the E-310 experiment ($Z_i = 1$, $n_0 = 1.12 \times 10^{17} \si{cm^{-3}}$, $\gamma = 19570$, $\sigma = 5\, \si{\mu m}$), $K \gg 1$ for the vast majority of particles in the beam and thus we are only interested in the wiggler regime.

In order to reconstruct the beam parameters from the betatron radiation signature measured by the spectrometer, we begin with a numerical model that takes beam parameters as an input and outputs the double differential radiation spectrum. Three such numerical models are discussed below. We then will use maximum likelihood estimation, expectation maximization, or machine learning to invert this model. Beam parameter reconstruction is analogous to the reconstruction of the radiation spectrum from the scintillator image described in the previous section.

\subsection{Model I: Idealized particle tracker with Liénard-Wiechert Radiation} \label{sec:model1}

We developed a betatron radiation code which tracks particles through idealized fields and computes radiation using Liénard–Wiechert potentials. The code is written in C++ and parallelized using MPI. First, macroparticles are randomly sampled from a beam distribution. Next, particles are tracked through idealized acceleration and focusing fields using a 4$^\mathrm{th}$ order Runge-Kutta method. Finally, the particle trajectories are used to numerically integrate the Liénard–Wiechert integral,

\begin{equation} \label{eq:lw}
 \bm{V}_i = \sqrt{\frac{e^2}{16 \pi^3 \epsilon_0 \hbar c}} \int_{t_i}^{t_f} \frac{\bm{n} \times \left(\left( \bm{n} - \bm{\beta} \right) \times \dot{\bm{\beta}} \right)}{\left(1 - \bm{n} \cdot \bm{\beta}\right)^2} e^{i \omega ( t - \bm{n} \cdot \bm{r}(t) / c )} \mathrm{d}t,
\end{equation}

\noindent where $\bm{n}$ is the normal vector in the direction of the radiation, $\bm{r}$ and $\bm{\beta}$ are the position and velocity over $c$ respectively which are computed by the particle tracking code, and $\omega$ is the angular frequency of the radiation. The radiation contribution from each particle is summed to give the double differential radiation spectrum, $\frac{d^2 I}{d\Omega d\epsilon} = \left| \sum_i \bm{V}_i \right| ^2$. The spectrum can be computed by numerically integrating this double differential over solid angle, $d\Omega$. If the beam is cylindrically symmetric in position and momentum space, while each particle's radiation is not cylindrically symmetric, the overall radiation generated by the beam can be evaluated over a 2D grid instead of a 3D grid which decreases computation time. While this simplification is often used, the code is still able to compute non-cylindrically symmetric radiation. Additionally, there are ample opportunities for parallelization of the code since the $\bm{V}_{i}$ add linearly and thus this computation is embarrassingly parallel.

\begin{figure}[h!]
    \centering
    \includegraphics[width=0.61\textwidth]{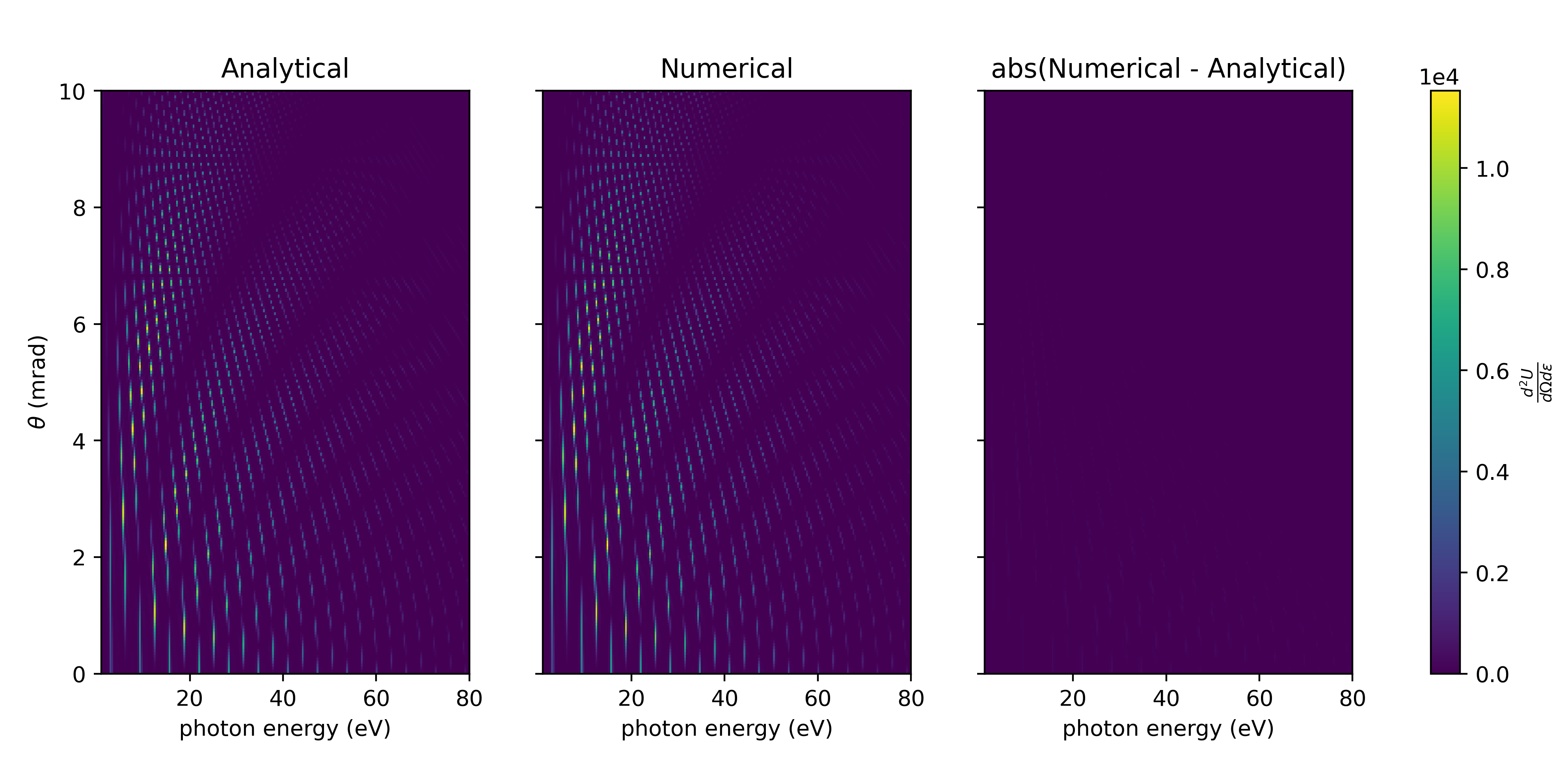}
    \includegraphics[width=0.4\textwidth]{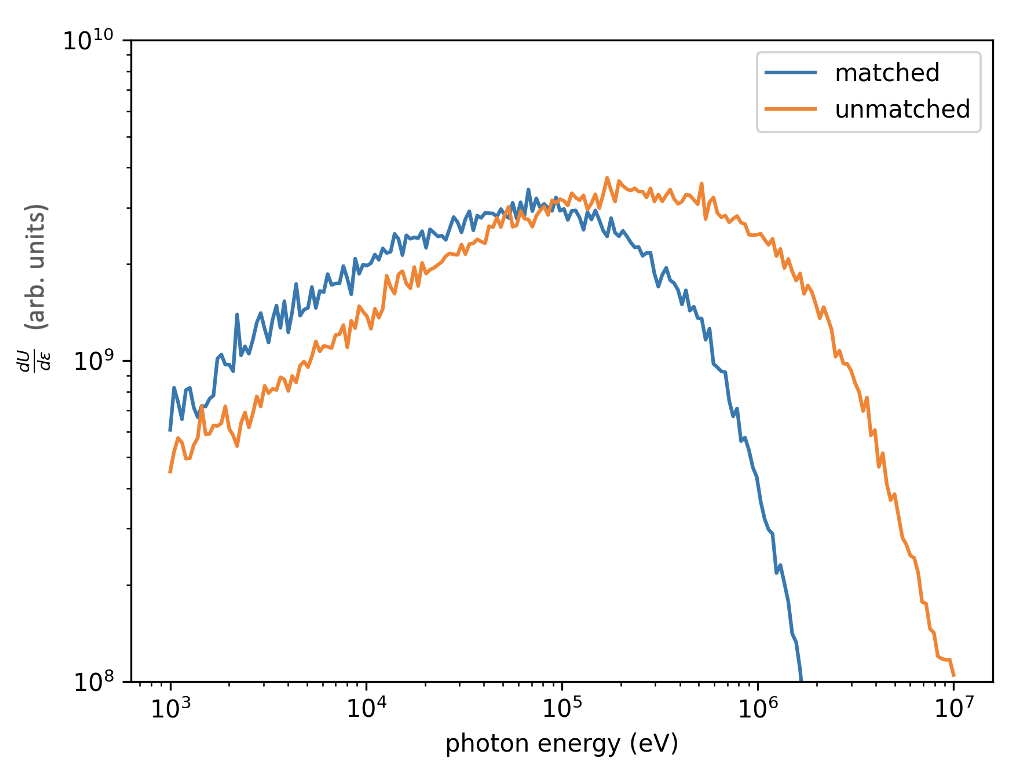}
    \caption{\textbf{Left:} Double differential spectrum of single particle, strong undulator radiation computed analytically and numerically. Radiation in this case is non-cylindrically symmetric and the radiation spectra plotted have azimuth $\phi = 0$. The absolute error between analytical and numerical results is also shown. The undulator has length $L = \SI{10}{cm}$, period $\lambda_u = \SI{1}{cm}$, and undulator parameter $K = 2$. The particle has energy $\SI{100}{MeV}$ and starts offset from the axis with no transverse velocity. \textbf{Right:} Radiation spectra for a matched and mismatched E-310 drive beam computed using Model I. Simulation parameters are shown in Table \ref{tab:E-310params}.}
    \label{fig:lwbenchmark}
\end{figure}

To validate this code, a simple benchmark is performed by comparing single particle, strong undulator radiation computed by the code with analytical expressions from \cite{hofmann}. A plot of the analytical and numerical double differential spectra and absolute error is shown in Fig.~\ref{fig:lwbenchmark}. This figure shows that the Liénard–Wiechert code correctly computes the strong undulator spectrum. This method was used to calculate the radiated spectra for both a matched and mismatched E-310 drive beam, with results shown in Fig.~\ref{fig:lwbenchmark}. The parameters of this simulation are shown in Table \ref{tab:E-310params}.

\subsection{Model II: Quasi-static particle-in-cell with Liénard-Wiechert radiation} \label{sec:model2}

Quasi-static PIC codes split the physics of the beam-plasma interaction into two timescales: the fast timescale of the plasma response, and the slow timescale of the beam evolution, significantly decreasing the computational cost of PWFA simulations. We use the quasi-Static code QuickPIC \cite{qpic1} to simulate relevant PFWA beam-plasma dynamics. While QuickPIC does not directly compute radiation, we modified it to output particle trajectories and input those trajectories into the Liénard–Wiechert code discussed in the previous section. A similar method was used in \cite{P_San}. Rather than compute radiation from every particle trajectory output by QuickPIC, a subset of those trajectories was randomly sampled and the result was scaled to the number of physical particles. A set of E-310 beam parameters and a resulting radiation spectrum calculated using this method are shown in Table \ref{tab:E-310params}. The shapes of the matched and unmatched spectra agree with the results computed by Model I in Fig.~\ref{fig:lwbenchmark}. Work on this model is ongoing and will be presented in a later paper. 

Parts A-D of Fig. 4 show the evolution of beam spot size and normalized emittance for unmatched beams in ramped and uniform plasmas simulated using QuickPIC.

\begin{table}[ht]
\begin{minipage}[b]{0.5\textwidth}
    \centering
    \begin{tabular}{|c|c|c|}
    \hline
    Parameter & Value & Unit \\
    \hline
    \multicolumn{3}{|c|}{Drive Beam} \\
    \hline
    $E$ & $10$ & $\si{GeV}$  \\
    $Q$ & $0.5$  & $\si{nC}$ \\
    $\sigma_{x, \text{matched}}$ & 0.89 & $\si{\mu m}$ \\
    $\sigma_{y, \text{matched}}$ & 0.89 & $\si{\mu m}$ \\
    $\sigma_{x, \text{unmatched}}$ & $4.5$ & $\si{\mu m}$ \\
    $\sigma_{y, \text{unmatched}} $ & $4.5$ & $\si{\mu m}$ \\
    $\sigma_z$ & $12.15$ & $\si{\mu m}$ \\
    $\epsilon_{n,x}$ & $5$ & $\si{\mu m }$ \\
    $\epsilon_{n,y}$ & $5$ & $\si{\mu m }$ \\
    \hline
    \end{tabular}
    \begin{tabular}{|c|c|c|}
    \hline
    \multicolumn{3}{|c|}{Plasma} \\
    \hline
    $n_0$ & $1.12 \times 10^{17}$ & $\si{cm^{-3}}$ \\
    $L$ & $60$ & $\si{cm}$ \\
    \hline
    \end{tabular}
    \end{minipage}\hfill
    \begin{minipage}[b]{0.5\textwidth}
    \centering
    \includegraphics[width=\textwidth]{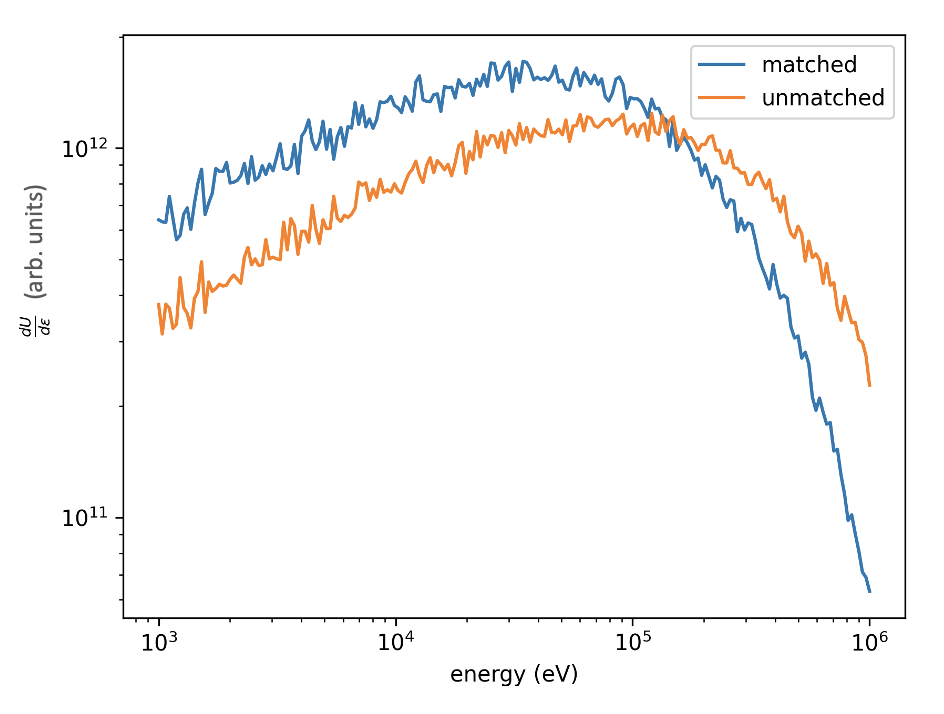}
    \caption{\textbf{Left}: E-310 drive beam and plasma parameters. \textbf{Right}: Radiation spectrum computed using Model II and these parameters.}
     \label{tab:E-310params}
    \end{minipage}
\end{table}

\begin{figure}[h!]
    \centering
    \includegraphics[width=\textwidth]{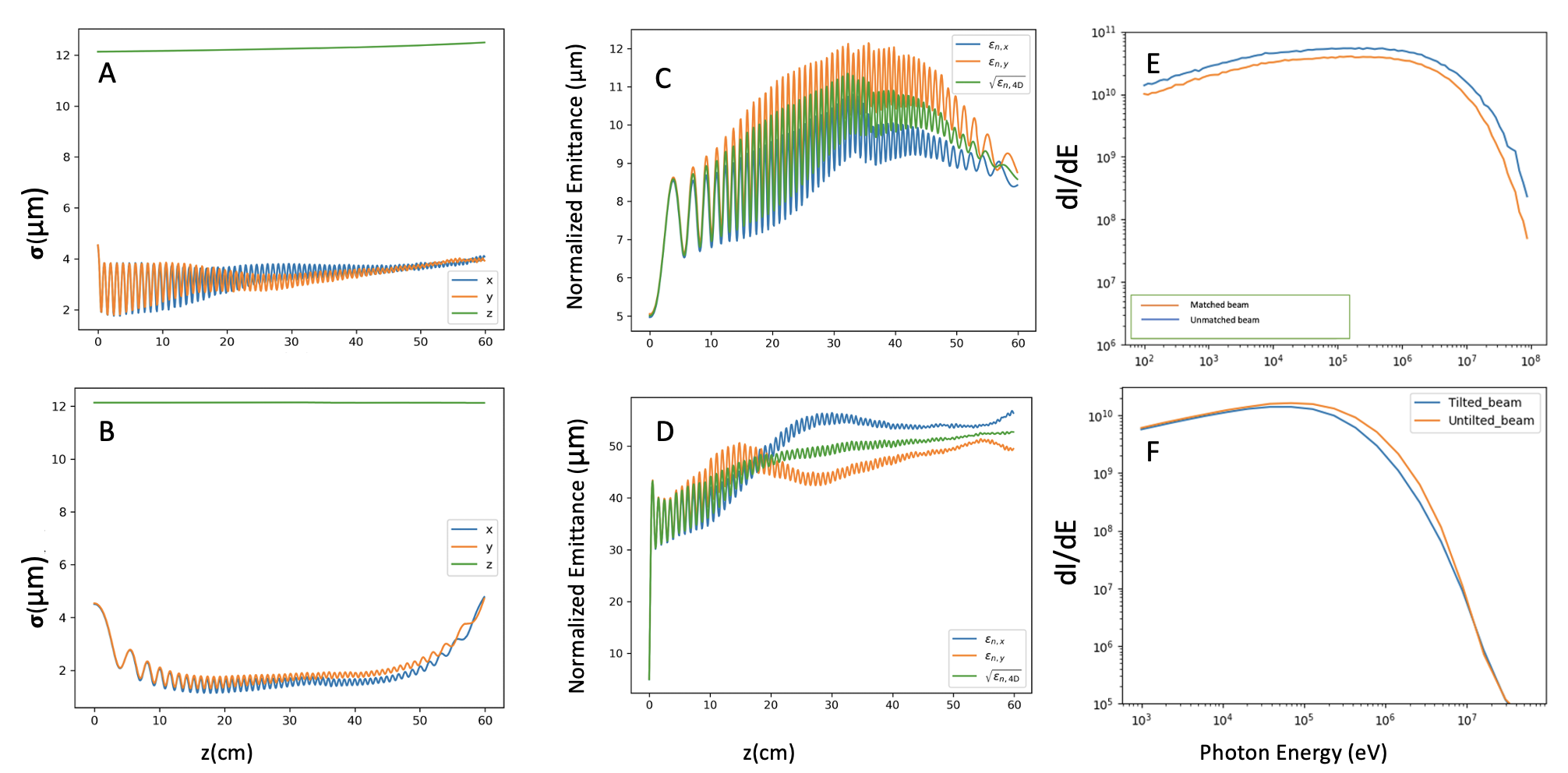}
    \caption{\textbf{(A-D)} Drive beam dynamics studies in QuickPIC, illustrating the use of a ramped profile to minimize emittance growth. \textbf{(A, C)} Spot sizes and emittances for an unmatched drive beam traveling through a ramped plasma: the plasma profile used a Gaussian upramp from $0-20\, \si{cm}$ with $\sigma = 20 / 3$ cm, followed by a uniform plasma density from $20-40$ cm, and then symmetric Gaussian downramp. \textbf{(B, D)} Spot sizes and emittances for an unmatched drive beam traveling through a uniform plasma. \textbf{(E, F)} The betatron radiation spectra from EPOCH simulations on \textbf{(E)} matched and unmatched beams and \textbf{(F)} tilted and untilted beams.
    }
    \label{fig:drivebeamstudies}
\end{figure}

\subsection{Model III: Full particle-in-cell with Monte Carlo QED Radiation} \label{sec:model3}

EPOCH is a 3D, fully explicit PIC code which uses a Monte Carlo QED model to simulate radiation generation \cite{epoch}. A primary challenge with 3D explicit codes is the artificial slow-down of the speed of light on a finite-difference (FDTD) grid. This means, for example, that a relativistic electron propagating along a straight line with constant velocity in free-space will unphysically emit numerical Cherenkov radiation (NCR) at wavelengths corresponding to the grid cell size, and may even grow as an instability by imprinting into the current profile \cite{Num_cherenkov}. In EPOCH, we use a dispersion-reduced FDTD solver \cite{lehe} and an 8-point, compensated \cite{J_Vay} linear current filter to mitigate this effect. Such schemes are imperfect and can slightly alter the Fourier content of fields at the grid resolution. However, the radiation model in EPOCH is photon-based, not field-based, and the emitted radiation wavelengths are well beyond the grid resolution. Therefore, we expect that there is minimal interference from the smoothing filter and microscopic details of the dispersion on our results, while retaining the benefit of smooth fields to be used in QED calculations. One of the primary challenges associated with using EPOCH is that large computational resources are needed to properly resolve physically relevant length-scales, especially in the matched beam case.

 Figure \ref{fig:drivebeamstudies} \textbf{E} compares the radiation spectra between a matched and unmatched beam in a ramped plasma. Figure \ref{fig:drivebeamstudies} \textbf{F} illustrates the difference in spectra between a matched beam and a tilted beam. Of particular note is that the photons in the tilted beam span an angular range of 6 mrad while the untilted beam spans only 1 mrad.

\section{CONCLUSION}

Betatron radiation is a powerful diagnostic to experimentally assess emittance growth, beam parameters, and the presence of instabilities, which are of key importance for the next generation of PWFA experiments. In this paper we have discussed the spectrometer currently under development at UCLA, algorithms for the reconstruction of radiation and beam parameters from the spectrometer output, and three models for betatron radiation from plasma accelerators. This work is of high significance to not just E-310 and other experiments at the FACET-II facility but also to future plasma acceleration experiments that utilize radiation diagnostics. Work is still ongoing to prepare for E-310 and other FACET-II experiments, and the results presented in this paper must be built upon before beam parameters can be reconstructed in an experimental context. On the hardware side, development and construction of the spectrometer is ongoing. On the computational side, the complexity of simulations must be increased to give a full picture of the relevant physics. In particular, witness beams must be included in the simulations, and in the case of Model III, the creation of the witness beam via laser ionization.


\section{ACKNOWLEDGMENTS}

This work was supported by DE-SC0009914 (UCLA) and the STFC Liverpool Centre for Doctoral Training on Data Intensive Science (LIV.DAT) under grant agreement ST/P006752/1. This work used computing resources provided by the STFC Scientific Computing Department’s SCARF cluster.

\nocite{*}
\bibliographystyle{aac}%
\bibliography{aac2020_latex}%

\end{document}